\documentclass[a4paper,11pt]{article}
\usepackage{jinstpub} % for details on the use of the package, please see the JINST-author-manual
\usepackage{lineno}
\usepackage{siunitx}
\usepackage{subcaption}

\title{\boldmath The H2M Monolithic Active Pixel Sensor \textemdash \ characterizing non-uniform in-pixel response in a 65~nm CMOS imaging technology}

\author[a,1]{S.~Ruiz~Daza%
\note{Corresponding author.},}
\author[b]{R.~Ballabriga,}
\author[c]{E.~Buschmann,}
\author[b]{M.~Campbell,}
\author[d]{R.~Casanova~Mohr,}
\author[b]{D.~Dannheim,}
\author[a]{J.~Dilg,}
\author[b]{A.~Dorda,}
\author[a]{F.~King,}
\author[a]{O.~Feyens,}
\author[b]{P.~Gadow,}
\author[a]{I.M.~Gregor,}
\author[a]{K.~Hansen,}
\author[a]{Y.~He,}
\author[a]{L.~Huth,}
\author[b]{I.~Kremastiotis,}
\author[b,2]{C.~Lemoine%
\note{Also at Université de Strasbourg, France.},}
\author[a]{S.~Maffessanti,}
\author[a]{L.~Mendes,}
\author[b]{Y.~Otarid,}
\author[a]{C.~Reckleben,}
\author[b]{S.~Rettie,}
\author[a]{M.A.~del~Rio~Viera,}
\author[a]{J.~Schlaadt,}
\author[a]{A.~Simancas,}
\author[b]{W.~Snoeys,}
\author[a]{S.~Spannagel,}
\author[a]{T.~Vanat,}
\author[a]{A.~Velyka,}
\author[a]{G.~Vignola,}
\author[a,3]{H.~Wennl{\"o}f%
\note{Now at Nikhef, Amsterdam.}}

\affiliation[a]{Deutsches Elektronen-Synchrotron DESY,\\
Notkestr. 85, 22607 Hamburg, Germany}

\affiliation[b]{CERN,\\
    Esplanade des Particules 1, Geneva, Switzerland}
   
\affiliation[c]{Brookhaven National Laboratory (BNL),\\
    New York 11973-5000,
   Upton, USA}

\affiliation[d]{
    Institut de Física d’Altes Energies (IFAE),\\
    Edifici CN, UAB campus, 08193 Bellaterra (Barcelona), Spain}

\emailAdd{sara.ruiz.daza@desy.de}

\abstract{The high energy physics community recently gained access to the TPSCo \SI{65}{\nano\meter} ISC (Image Sensor CMOS), which enables a higher in-pixel logic density in monolithic active pixel sensors (MAPS) compared to processes with larger feature sizes. 
To explore this novel technology, the Hybrid-to-Monolithic (H2M) test chip has been designed and manufactured. 
The design follows a digital-on-top design workflow and ports a hybrid pixel-detector architecture, with digital pulse processing in each pixel, into a monolithic chip. 
The chip matrix consists of 64$\times$16 square pixels with a size of \SI{35}{}$\times$\SI{35}{\micro\meter^2}, and a total active area of approximately \SI{1.25}{\meter\meter^2}. 
The chip is operated and read out using the Caribou DAQ system. The measured threshold dispersion and noise agree with the expectation from front-end simulations.
However, a non-uniform in-pixel response related to the size and location of the n-wells in the analog circuitry has been observed in test beam measurements and will be discussed in this contribution.
This asymmetry in the pixel response, enhanced by the \SI{35}{\micro\meter} pixel pitch \textemdash \ larger than in other prototypes \textemdash \ and certain features of the readout circuit, has not been observed in prototypes with smaller pixel pitches in this technology.
}


\notoc 
\begin{document}
\maketitle 
\flushbottom
\newpage

\section{Introduction}
\label{sec:introduction}
Monolithic CMOS sensors enable the development of detectors with a low material budget and a low fabrication cost compared to hybrid pixels. By employing a small collection electrode, these sensors achieve small input capacitance, low analog power consumption, and an improved signal-to-noise ratio compared to sensors with larger collection electrodes. 
The availability of a \SI{65}{\nano\meter} CMOS imaging process to the high-energy physics community further enhances the potential of monolithic active pixel sensors (MAPS), as it allows for a higher density of in-pixel logic compared to processes with larger feature sizes. This \SI{65}{\nano\meter} process has been previously explored in prototypes such as the APTS~\cite{apts} and DPTS~\cite{dpts}.

The H2M (Hybrid-to-Monolithic) test chip has been developed to demonstrate the capabilities of the \SI{65}{\nano\meter} CMOS imaging process in tracking applications for future high-energy lepton experiments (such as CLIC or FCCee). The architecture of this monolithic chip is based on a hybrid readout chip (from the Timepix family), with the goal of learning about the challenges and potential associated with the porting process. The development aimed to design and evaluate a set of digital logic building blocks, a so-called digital cell library, with a smaller footprint (reduction of $\sim$25\% in logic area compared to other available standard cell libraries), as well as complex and fast in-pixel readout circuitry explained below.

\section{The H2M test chip}
\label{sec:h2m}

\begin{figure}[b]
\centering
\includegraphics[width=1\textwidth]{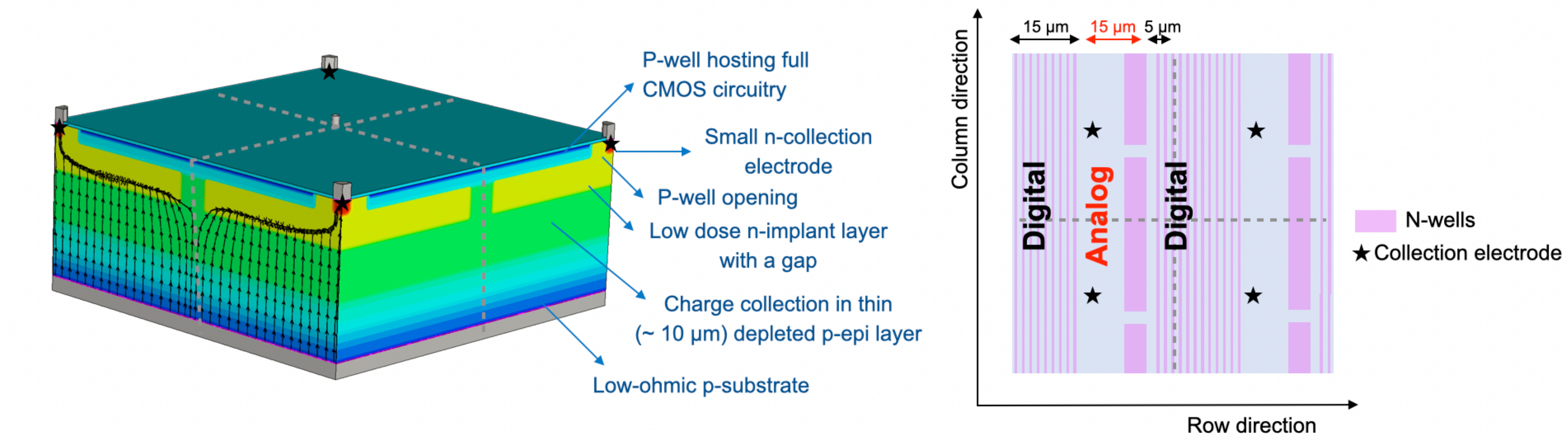}
\caption{Schematic of the sensor layout on the left, including the charge collection path. The position of the digital and analog circuitry hosted within the deep p-wells, as well as the position of n-wells therein, is schematically shown from a top view on the right. The pixel cell boundary is marked with a dashed gray line, and the n-collection electrodes with a star.} 
\label{fig:tcad}
\end{figure}

The H2M sensor is manufactured using a modified TPSCo \SI{65}{\nano\meter} ISC (Image Sensor CMOS)~\cite{walter}.
It comprises a small n-type collection electrode placed on a high-resistivity p-type epitaxial layer, which is grown on a low-resistivity substrate.
To enhance charge collection in the sensitive layer with a thickness of \SI{\sim10}{\micro\meter}, process modifications have been implemented, including a low-dose n-type implant with a gap at the pixel boundaries~\cite{walter}. The wafers are backside-thinned to \SI{50}{\micro\meter} physical thickness. A schematic of the sensor layout is illustrated in figure~\ref{fig:tcad}.

The chip matrix consists of 64$\times$16 square pixels with a size of \SI{35}{}$\times$\SI{35}{\micro\meter^2}, resulting in a total active area of approximately \SI{1.25}{\meter\meter^2}. 
Each pixel includes an analog front end that comprises a charge-sensitive amplifier (CSA) with Krummenacher feedback~\cite{krummenacher} followed by a discriminator. The design ensures a constant slope of the falling edge of the CSA output, adjustable through the feedback current (ikrum), where a lower ikrum effectively increases the integration time, enhancing the signal and improving the signal efficiency for a given threshold. 
The global threshold of the discriminator is selected with an 8-bit DAC, and threshold mismatch is compensated using a 4-bit threshold-tuning DAC per pixel. Individual pixels can be masked, and if needed, test pulses of varying amplitudes can be injected at the input stage of the CSA.
The digital logic processes the output of the discriminator. Both the CMOS analog and digital front ends are hosted within the deep p-wells. The analog front-end begins at the collection electrode and extends \SI{15}{\micro\meter} in the row direction, with the remaining space occupied by the digital readout circuitry. PMOS devices are shielded in n-wells inside the deep p-well. 

The chip can operate in four non-simultaneous acquisition modes: 8-bit Time-over-Threshold (ToT) for energy measurements, 8-bit Time-of-Arrival (ToA) with \SI{10}{\nano\second} binning for time measurements, photon counting, and triggered mode. 
ToA represents the time of the threshold crossings as the charge is collected, while the ToT is the duration of the signal above the threshold. 
Photon counting mode counts hits above the threshold. In triggered mode, a binary readout occurs after hit validation by an external trigger signal, using an 8-bit delay counter to accommodate the trigger delay.
The readout is integrated into the Caribou DAQ system~\cite{caribou}. It uses a \SI{40}{\mega\hertz} clock and is frame-based without zero suppression. 

The chip is fully functional, and it has been calibrated using radioactive sources. A single-pixel noise of \SI{33}{}~electrons r.m.s. and a threshold dispersion of the equalized matrix of \SI{45}{}~electrons have been measured, which agrees with the expectations from the front-end simulations.

\section{Performance in test beam}
\label{sec:testbeam}
\begin{figure}
  \begin{subfigure}{0.48\textwidth}
    \includegraphics[width=\linewidth]{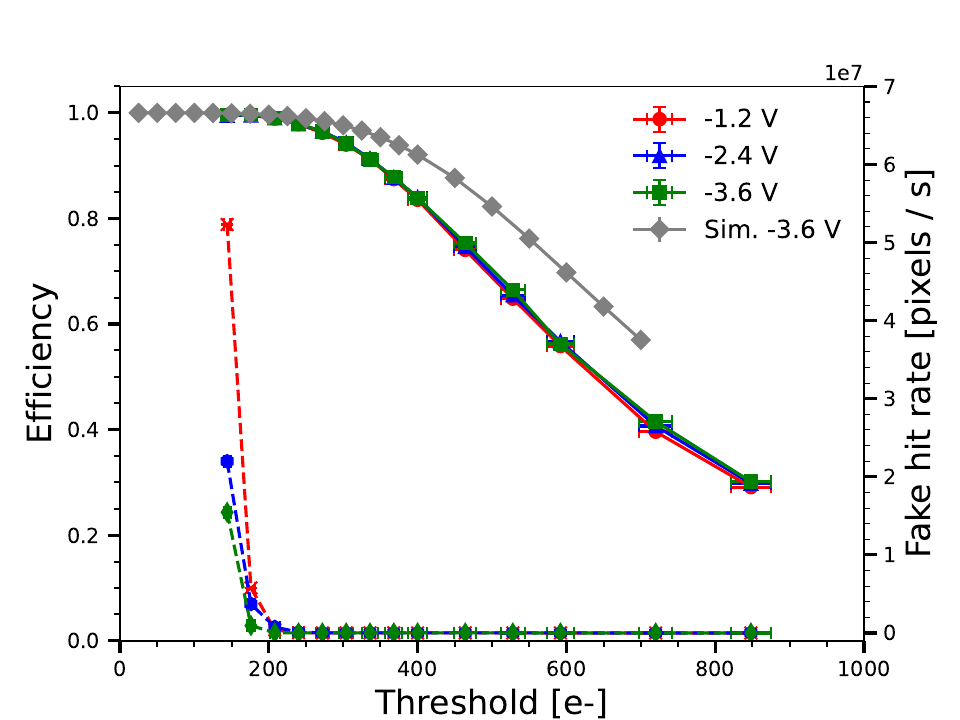}
    \caption{} \label{fig:eff_thl}
  \end{subfigure}%
  \hspace*{\fill}   % maximize separation between the subfigures
  \begin{subfigure}{0.48\textwidth}
    \includegraphics[width=\linewidth]{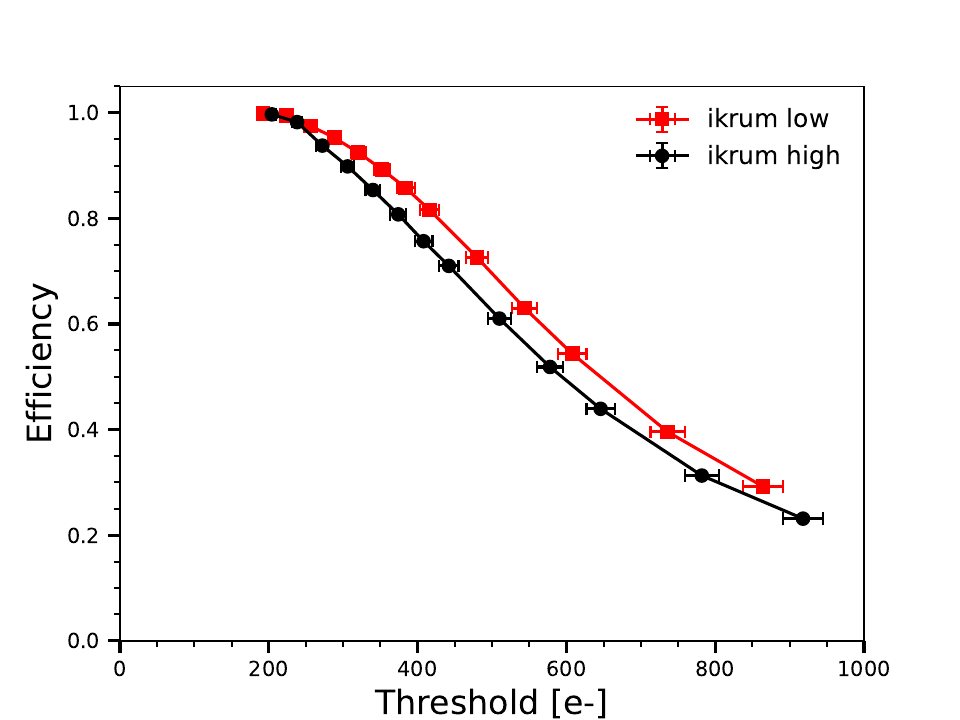}
    \caption{} \label{fig:ikrum_thl}
  \end{subfigure}%
\caption{(a) Measured efficiency (left axis) in solid lines and fake hit rate (right axis) in dashed lines as a function of the hit detection threshold for three different sensor bias voltages in triggered mode. The expected efficiency at \SI{-3.6}{V} obtained from simulations with generic profiles is also shown.
(b) Measured efficiency as a function of the hit detection threshold for two different feedback currents (ikrum) in ToA mode. The sensor is biased at  \SI{-1.2}{V}.} \label{fig:thl_scan}
\end{figure}
To study the performance of H2M in terms of particle detection, the chip has been tested at the DESY~II test beam facility~\cite{testbeam}, using the ADENIUM telescope~\cite{adenium} for particle tracking, and the Telepix2 detector serving as reference trigger and timing detector~\cite{telepix2}. For data analysis, the Corryvreckan framework is used~\cite{corry}. 

Figure~\ref{fig:eff_thl} shows the efficiency and fake hit rate as a function of the hit detection threshold for three different bias voltages measured in triggered mode with a fixed frame duration of \SI{500}{\nano\second} per event.
The bias voltage on the p-well/substrate influences the depletion of the sensor volume. At low bias voltages ($< \SI{2}{V}$), the area surrounding the collection electrode is not fully depleted, which increases the sensor capacitance and, consequently, noise and fake hit rate. 
Allpix$^2$ simulations, using the results from Technology computer-aided design (TCAD) simulations based on generic profiles~\cite{haakan}, are performed to obtain the efficiency as a function of the detection threshold, which is also included in figure~\ref{fig:eff_thl}. The difference between the measured and simulated efficiency can be explained by a non-uniformity in the in-pixel response, which will be further discussed below.

\begin{figure}
  \begin{subfigure}{0.33\textwidth}
    \includegraphics[width=\linewidth, trim={0.5cm 2cm 1cm 2cm},clip]{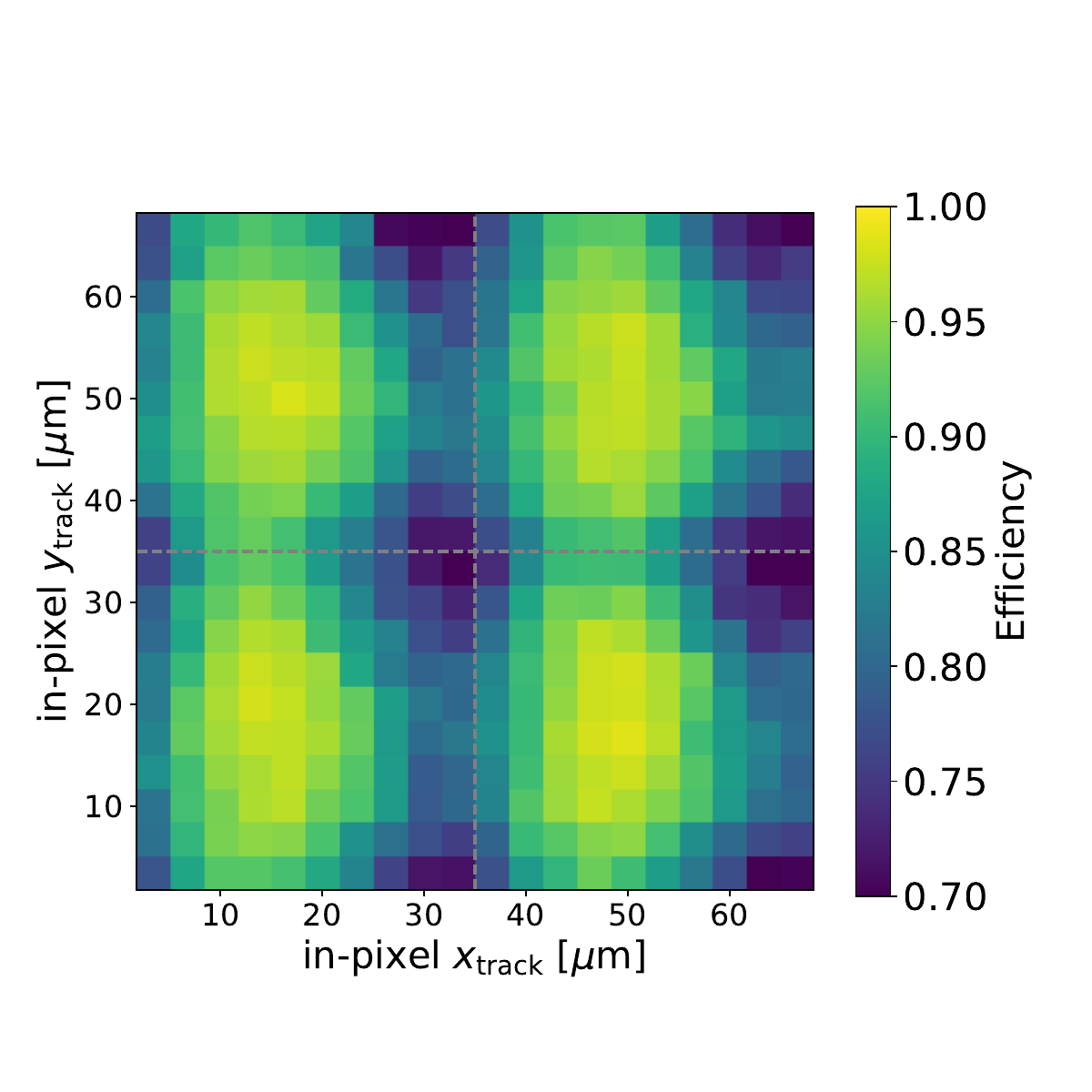}
    \caption{} \label{fig:eff_map}
  \end{subfigure}%
  \hspace*{\fill}   % maximize separation between the subfigures
  \begin{subfigure}{0.33\textwidth}
    \includegraphics[width=\linewidth, trim={0.5cm 2cm 1cm 2cm},clip]{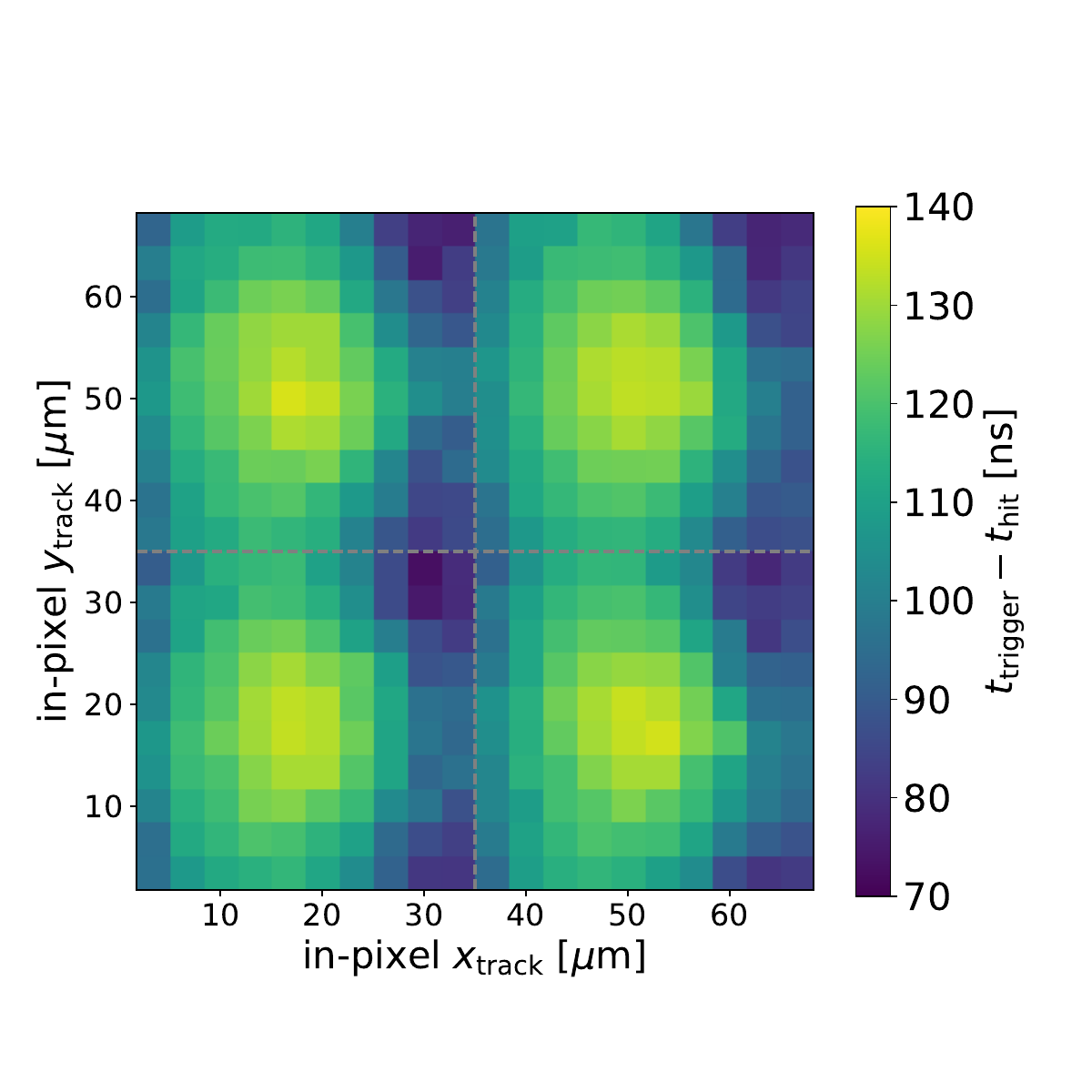}
    \caption{} \label{fig:toa_map}
  \end{subfigure}%
  \hspace*{\fill}
  \begin{subfigure}{0.33\textwidth}
    \includegraphics[width=\linewidth, trim={0.5cm 2cm 1cm 2cm},clip]{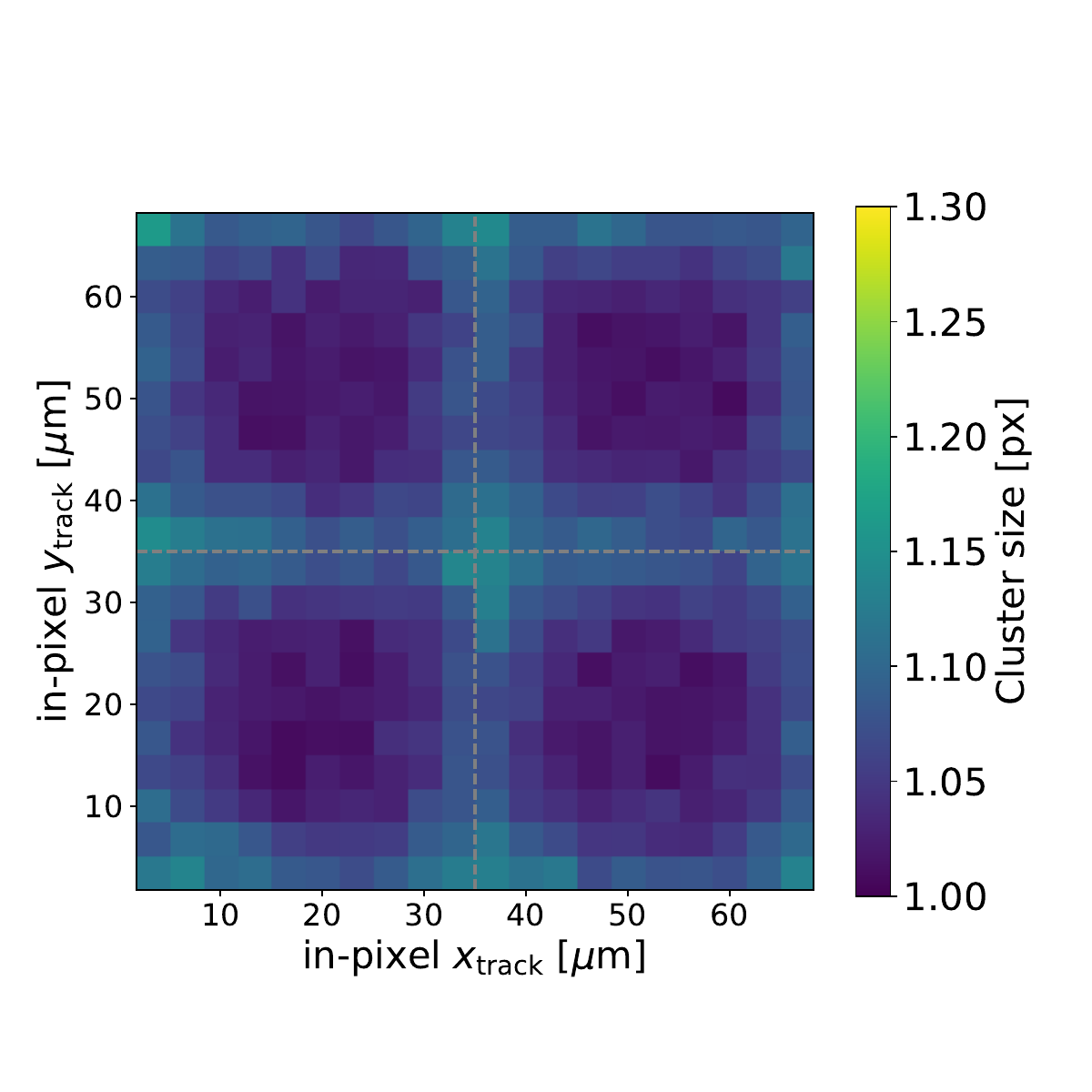}
    \caption{} \label{fig:clustersize}
  \end{subfigure}%
\caption{Measured efficiency (a), ToA (b), and cluster size (c) maps projected onto four pixels. In all figures, the sensor is biased at \SI{-1.2}{V}, and the hit detection threshold is \SI{332}{electrons}. The pixel cell boundary is marked with a dashed gray line.} \label{fig:inpixel}
\end{figure}

Figure~\ref{fig:eff_map} and~\ref{fig:toa_map} show the efficiency and ToA as a function of the in-pixel hit position. These maps are obtained by projecting all track intercepts into four pixels.
High efficiency and fast charge carrier collection occur near the collection electrode but decrease asymmetrically towards the edges and corners. 
Particularly, the drop in charge collection speed, resulting in reduced efficiency, is located beneath the n-well of the analog front-end (see figure~\ref{fig:tcad}). 
While the n-wells in the digital readout circuitry are thin and uniformly distributed, the n-well in the analog front-end has a width of about \SI{4}{\micro\meter}, which affects the electric field at the interface of the p-well and the low-dose n-implant layer, creating local potential wells that slow down charge collection along their paths (see figure~\ref{fig:tcad}). 
This does not influence charge sharing, and the cluster size map is therefore symmetric, as shown in figure~\ref{fig:clustersize}.

The simulations presented in figure~\ref{fig:eff_thl} do not consider any well structure within the deep p-wells. In order to fully understand and reproduce the measurements presented here, more realistic simulations, including the well of the analog front-end circuitry and its electronics simulations, have been performed. Those studies are summarized in~\cite{corentin}.

\begin{figure}[htbp]
\centering
\includegraphics[width=.55\textwidth]{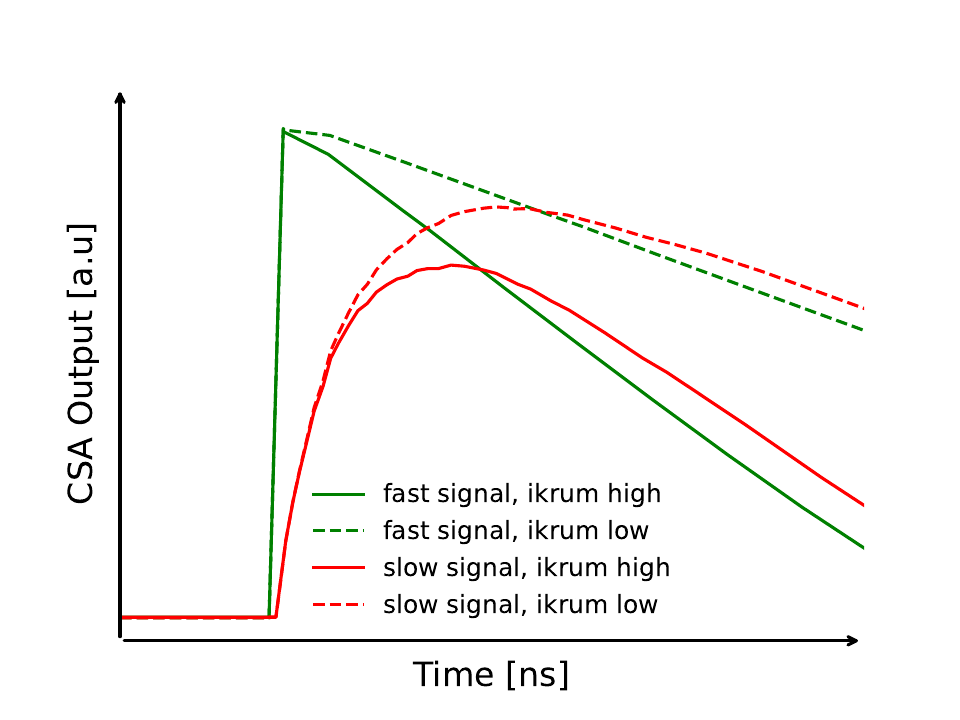}
\caption{Sketch showing the different impact of the ballistic deficit on slow and fast signals.}
\label{fig:sketch_ballistic_deficit}
\end{figure}

Figure~\ref{fig:sketch_ballistic_deficit} illustrates the CSA output for slow and fast signals for large and small ikrum. The loss of the CSA output signal height for slow signals occurs due to a mismatch between the charge collection time and the CSA response time. This effect is known as ballistic deficit. While the signal amplitude for the fast signals barely depends on ikrum, slow signals yield a different amplitude.
As a result, a low ikrum improves the overall chip efficiency, as demonstrated in figure~\ref{fig:ikrum_thl}. 
At low hit detection thresholds (< \SI{180}{electrons}) and larger bias voltages, the in-pixel response remains uniform, achieving an efficiency of \SI{99.6}{\%} at a threshold of \SI{144}{electrons} with the sensor biased at \SI{-3.6}{V}. 

This effect is enhanced by the fast analog front-end and the large pitch of H2M~\cite{corentin}. The integration time of a few nanoseconds contributes to the ballistic deficit by reducing the amplitude of slower signals compared to faster ones. 
Additionally, the relatively large pitch of \SI{35}{\micro\meter} results in significant spacing between the lateral electric field components of the collection electrodes and the gap, which makes it more sensitive to field perturbations due to the n-wells in the p-well region. This makes H2M the only chip in this technology where such non-uniform in-pixel responses have been observed.

\section{Conclusion} \label{sec:conclusion}
The H2M test chip, manufactured using a modified \SI{65}{\nano\meter} CIS, successfully ported a hybrid architecture with ToA and ToT counters in each pixel into a MAPS. 
Both the analog and the digital front-ends are fully functional. A single-pixel noise of \SI{33}{}~electrons r.m.s. and a threshold dispersion of \SI{45}{}~electrons for the equalized matrix has been measured, in agreement with predictions from front-end simulations.
Test beam measurements showed a non-uniform in pixel response, with efficiency and timing strongly depending on the track impact position.
This effect, amplified by the large pitch and fast front-end, has been correlated with the size and location of the n-wells of the analog circuitry. 
When operating the chip with high bias voltages and low ikrum, high efficiency, and uniform in-pixel response are achieved.
Complementary measurements using laser and radioactive sources, along with simulations, have been conducted to better understand the influence of n-wells on charge collection and to prevent the issue from appearing in future chip submissions.

\acknowledgments
The measurements leading to these results have been performed at the Test Beam Facility at DESY Hamburg (Germany), a member of the Helmholtz Association (HGF).\\
The developments presented in this contribution are performed in collaboration with the CERN EP R\&D programme on technologies for future experiments.\\
This project has received funding from the European Union’s Horizon 2020 Research and Innovation programme under GA no 101004761.

% Bibliography

%% [A] Recommended: using JHEP.bst file
%% \bibliographystyle{JHEP}
%% \bibliography{biblio.bib}

%% or
%% [B] Manual formatting (see below)
%% (i) We suggest to always provide author, title and journal data or doi:
%% in short all the informations that clearly identify a document.
%% (ii) please avoid comments such as "For a review'', "For some examples",
%% "and references therein" or move them in the text. In general, please leave only references in the bibliography and move all
%% accessory text in footnotes.
%% (iii) Also, please have only one work for each \bibitem.

\end{document}